\documentclass[aps,prc,twocolumn,groupedaddress,showpacs,floatfix]{revtex4}
\usepackage{epsfig}
\usepackage{bm}
\def\NEG#1{{\rlap/#1}}
\begin{document}

\preprint{NT@UW-2001-29}
\title{Pion-only, chiral  light-front model of the deuteron}
\author{Jason R.~Cooke}
\author{Gerald A.~Miller}
\affiliation{
Department of Physics \\
University of Washington \\
Box 351560 \\
Seattle WA 98195-1560}
\date{\today{}}

\begin{abstract}
We investigate the symptoms of broken rotational invariance, caused by
the use of light front dynamics, for deuterons obtained  using one- and
two-pion-exchange potentials. A large mass splitting between
states with $m=0$ and $m=1$ is found for the deuteron obtained from  the
one-pion-exchange (OPE) potential. The size of the splitting is smaller
when the chiral two-pion-exchange (TPE) potential is used. When the TPE
potential constructed without chiral symmetry is used, the deuteron
becomes unbound. These results arise from significant relativistic
effects which are much larger than those of  the  Wick-Cutkosky model
because of the presence of the tensor force. 
\end{abstract}


\maketitle

\section{Introduction} \label{ch:int}

Recently, there has been much interest in effective field theories (EFT)
for nuclear physics. In particular, the expansion of the nuclear forces
about the chiral limit gives good results
\cite{Weinberg:um,Ordonez:1995rz}. Attempts have been made to apply
pion-less EFTs to study deuteron properties 
\cite{Kaplan:1998tg,vanKolck:1998bw,Chen:1999tn}. However, it has been
shown that pions are required to obtain a well-controlled theory
\cite{Beane:2001bc}. 

It is not surprising that the pions are necessary. Indeed, the gross
features of the deuteron are determined principally by the
one-pion-exchange (OPE) potential \cite{Ericson}, a consequence of
EFT. However, the short-distance behavior of the na\"{\i}ve OPE potential is
singular, and must be regulated. Physically, this regulation is provided
by the exchange of more massive mesons and multi-meson exchanges. EFT
tells us that these high-energy parts can be replaced by anything that
has the correct form dictated by symmetry, and the low energy properties
of the deuteron will be left unaffected. 

The expectation that pionic effects are very important is related to
work by Friar, Gibson, and Payne (FGP)\cite{Friar:1984wi} who obtained
non-relativistic deuteron wave functions using only the OPE
potential. In their  calculations a pion-nucleon form factor 
replaces the high-energy physics. The only
influence of chiral symmetry in the FGP
model is to require that that the  pion-nucleon
coupling be of the $\bm{\tau}\bm{\sigma}\cdot\bm{\nabla}$ form.

In this paper, we build upon the FGP model by performing a relativistic
calculation using light-front dynamics. In addition, we go beyond the
OPE potential and include the two-pion-exchange (TPE) potential. Chiral
symmetry is known to have profound implications for the TPE interaction,
so that we will be able to study the effect that ignoring chiral
symmetry has on the deuteron.

Our plan is to  introduce a model Lagrangian for nuclear physics using
only nucleons and pions. A Lagrangian which includes chiral symmetry
\cite{Miller:1997cr} has been used to compute the OPE and TPE
potentials for a new light-front nucleon-nucleon potential
\cite{Miller:1997cr,Cooke:2001dc,Cooke:2001kz} which involves the
exchange six different mesons. We retain only the contributions arising
from pionic exchanges here in Sect. \ref{ch:pionly}. 
The resulting  pion-only
potentials are then used to see if the deuteron state is bound and if
so, to compute  the binding energy in Sect.~\ref{ch:results}.
A final section \ref{ch:conclusions} is
devoted to presenting a brief set of conclusions.

\section{Model} \label{ch:pionly}

We consider a pion-only light-front nucleon-nucleon potential derived
from a nuclear Lagrangian. This model is inspired by the
non-relativistic one-pion-exchange model used by Friar, Gibson, and
Payne \cite{Friar:1984wi}. We generalize their model to form the basis
of our pion-only light-front model, which includes relativity
automatically. This pion-only model is essentially the same as the model
presented in Ref.~\cite{Cooke:2001kz} restricted to pions only.

Our starting point is a nuclear Lagrangian
\cite{Miller:1997cr, Miller:1997xh,Cooke:2001kz, Miller:1999ap}
which incorporates a non-linear chiral
model for the pions. The Lagrangian is based on the linear
representations of chiral symmetry used by Gursey
\cite{Gursey:1960yy}. It is invariant (in the limit where
$m_\pi\rightarrow0$) under chiral transformations.

The pion-only model prescribes the use of nucleons $\psi$
and the $\pi$ meson, which is a pseudoscalar isovector. The Lagrangian
${\mathcal L}$ is given by
\begin{eqnarray}
{\mathcal L} &=&
 \frac{1}{4}f^2 \mbox{Tr} (\partial_\mu U \, \partial^\mu U^\dagger)
+\frac{1}{4}m_\pi^2f^2 \, \mbox{Tr}(U +U^\dagger-2)
\nonumber\\&&
+\overline{\psi} \Big[ i\NEG\partial - U M \Big] \psi \label{eq:mainNNlag},
\end{eqnarray}
where the bare mass of the nucleon is $M$ and the pion is $m_\pi$. The
unitary matrix $U$ can be chosen to have one of the three forms $U_i$:
\begin{eqnarray}
U_1 &\equiv& e^{i  \gamma_5 \bm{\tau\cdot\pi}/f}, \\ 
U_2 &\equiv& \frac
{1+i\gamma_5\bm{\tau}\cdot\bm{\pi}/2f}
{1-i\gamma_5\bm{\tau}\cdot\bm{\pi}/2f}, \\ 
U_3 &\equiv& \sqrt{1-\pi^2/f^2}+i\gamma_5\bm{\tau\cdot\pi}/f, \label{us}
\end{eqnarray}
which correspond to different definitions of the fields. Note that
each of these definitions can be expanded to give
\begin{eqnarray}
U &=& 1 + i \gamma_5 \frac{\bm{\tau}\cdot\bm{\pi}}{f}
- \frac{\pi^2}{2f^2} + {\mathcal O}
\left(\frac{\pi^3}{f^3}\right) \label{pi:pertU}.
\end{eqnarray}
In this paper, we consider at most two pion exchange potentials, so we
consider $U$ to be defined by Eq.~(\ref{pi:pertU}).

In the limit where $m_\pi\rightarrow0$, the Lagrangian in
Eq.~(\ref{eq:mainNNlag}) is invariant under the chiral transformation
\begin{eqnarray}
\psi \rightarrow e^{i \gamma_5 \bm{\tau}\cdot\bm{a}}\psi, \qquad
U \rightarrow e^{-i \gamma_5 \bm{\tau}\cdot\bm{ a}} \;U\; 
e^{-i \gamma_5 \bm{\tau}\cdot \bm{a}}.
\label{chiral}\end{eqnarray}

We use the Lagrangian to obtain the OPE and TPE light-front
nucleon-nucleon potentials
\cite{Miller:1997xh,Miller:1997cr,Miller:1999ap,Cooke:2001kz,Cooke:2001dc}.
We find that there are three classes of TPE potentials. The first class,
the TPE box diagrams, consists of diagrams where the pion lines do not
cross. The second class, the TPE contact diagrams, consists of diagrams
where at least one of the vertices is a two-pion contact vertex, as
demanded by chiral symmetry. The third class is the TPE crossed diagrams,
where the pion lines cross.

As discussed in
Refs.~\cite{Cooke:1999yi,Cooke:2000ef,Cooke:2001kz,Cooke:2001dc}, we have
some freedom in deciding which TPE diagrams to include in our
potential. In particular, we may neglect just the crossed diagrams, or
both the crossed and contact diagrams. Although neglecting these
diagrams may affect the exact binding energy calculated, we should find
a partial restoration of rotational invariance as compared to using just
the OPE potential. 

We also want to keep the potentials chirally symmetric as well. Whereas
rotational invariance of the potential is partially restored by including
higher-order potentials, chiral symmetry is restored by including 
the $\pi\pi$ contact interaction graphs to the same order as the
non-contact graphs.

\begin{figure}
\begin{center}
\epsfig{angle=0,width=3.0in,height=0.75in,file=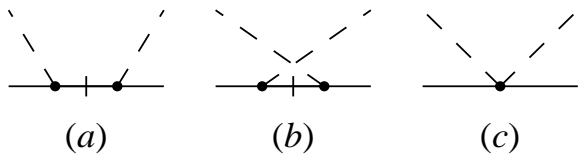}
\caption{The non-vanishing diagrams for pion-nucleon scattering at
threshold: (a) ${\mathcal M}_U$, (b) ${\mathcal M}_X$, and (c)
${\mathcal M}_C$.
\label{fig:nt.chiralcan}}
\end{center}
\end{figure}

Chiral symmetry tells us that for pion-nucleon scattering at threshold,
the time-ordered graphs approximately cancel
\cite{Miller:1997cr}. Furthermore, upon closer examination, we find that
all the light-front time-ordered graphs for the scattering amplitude
vanish except for the two graphs with instantaneous nucleons and the
contact graph. These graphs are shown in
Fig.~\ref{fig:nt.chiralcan}. Using the Feynman rules
Refs.~\cite{Miller:1997cr, Miller:1997xh, Miller:1999ap,Cooke:2001kz,Cooke:2001dc}, 
and denoting the
nucleon momentum by $k$ and the pion momentum by $q$, we find that
\begin{eqnarray}
{\mathcal M}_U &=& C \frac{\tau_i\tau_j}{2(k^++q^+)}u(k')\gamma^+u(k), \\
{\mathcal M}_X &=& C \frac{\tau_j\tau_i}{2(k^+-q^+)}u(k')\gamma^+u(k), \\
{\mathcal M}_C &=& C \frac{-\delta_{i,j}}{M}        u(k')        u(k),
\end{eqnarray}
where the factors common to all amplitudes are denoted by $C$.

For threshold scattering, we take $k^+=M$ and $q^+=m_\pi$. In that
limit, we find
\begin{eqnarray}
{\mathcal M}_U &=& C'
\frac{ \delta_{i,j}+i\epsilon_{i,j,k}\tau_k}{2(M+m_\pi)},\\
{\mathcal M}_X &=& C'
\frac{ \delta_{i,j}-i\epsilon_{i,j,k}\tau_k}{2(M-m_\pi)},\\
{\mathcal M}_C &=& C'\frac{-\delta_{i,j}}{M},     
\end{eqnarray}
where $C'=C \overline{u}(k')u(k)$. In the limit that $m_\pi\rightarrow
0$, the sum of these three terms vanishes. The term in these equations
proportional to $\tau_k$ is the famous Weinberg-Tomazowa term
\cite{Brandsden:1973,Adler:1968}.

The fact that the amplitudes cancel only when the contact interaction is
included demonstrates that chiral symmetry can have a significant effect
on calculations. In terms of two-pion-exchange potentials, this result
means that the contact potentials cancel strongly with both the iterated
box potentials and the crossed potentials. This serves to reduce the
strength of the total two-pion-exchange potential, which should lead to
more stable results.

However, since we do not use the crossed graphs for the nucleon-nucleon
potential, we must come up with a prescription which divides the contact
interactions into two parts which cancel the box and crossed diagrams
separately. We do this by formally defining two new contact
interactions, so that 
\begin{eqnarray}
{\mathcal M}_{C_U} &\equiv& \frac{M}{2(M+\lambda m_\pi)} {\mathcal M}_C,
\label{csa1} \\
{\mathcal M}_{C_X} &\equiv& {\mathcal M}_C - {\mathcal M}_{C_U},
\label{csa2}
\end{eqnarray}
and the value of $\lambda$ is of order unity.
With these definitions, we find that at threshold and in the chiral
limit,
\begin{eqnarray}
{\mathcal M}_U + {\mathcal M}_{C_U} &=& 0, \\
{\mathcal M}_X + {\mathcal M}_{C_X} &=& 0.
\end{eqnarray}
Since the prescription given by equations (\ref{csa1}) and (\ref{csa2}) leads to
the correct results in the chiral limit for pion-nucleon scattering, we use it
to factor the TPE potentials so that chiral symmetry can be approximately
maintained while neglecting the crossed potentials. 

We find that the results of our TPE calculation which include approximate chiral
symmetry have a very weak dependence on the value of $\lambda$, provided
that $0<\lambda<1$. As such, we show only the results for
$\lambda=1$. 
These features indicate that we can incorporate approximate
chiral symmetry without including crossed graphs simply by weighting
each contact interaction graph with a factor of
$\frac{M}{2(M+m_\pi)}$.

\section{Results} \label{ch:results}

To numerically calculate the bound states for these pion-only
potentials, we must choose values for the potential parameters.
As a starting point, we look for inspiration from the non-relativistic
one-pion-exchange model used by Friar, Gibson, and Payne (FGP)
\cite{Friar:1984wi}. The basic parameters of the model are the mass of
the nucleon ($M=938.958$~MeV), the mass of the pion
($m_\pi=138.03$~MeV), and the pion decay constant
($f_0^2=0.079$), which correlates to a coupling constant of 
$\frac{g_\pi^2}{4\pi}=\frac{4M^2f_0^2}{m_\pi^2}=14.6228$ in our
formalism. 

The FGP model uses family of $n$-pole pion-nucleon form factors, with the form
\begin{eqnarray}
F(q)&=&\left( \frac
{\Lambda^2-m_\pi^2}
{\Lambda^2+\bm{q}^2} \right)^n,
\end{eqnarray}
and the parameter $\Lambda$ is fit to reproduce the deuteron binding energy for
a given value of $n$. 

We can obtain the FGP model from our light-front nucleon-nucleon
potential by considering only the OPE potential and performing a
non-relativistic reduction. To understand our results, it is important
to first consider how well the light-front potential and wave functions
approximate the non-relativistic potential and wave functions. 

It is useful to start by reviewing some of the properties of the
deuteron. First, we note that the deuteron is very lightly
bound. Second, although a majority of the deuteron wave function resides
in the non-relativistic regime, it has a high-momentum tail that falls
off rather slowly, as $\frac{1}{k^4}$ \cite{Machleidt:1987hj}. Recalling
our experience with the Wick-Cutkosky model \cite{Cooke:1999yi}, where
we found that mass splitting between states with different $m$ values is
small for lightly bound states, it may appear plausible that masses of
the $m=0$ and $m=1$ states of the deuteron are approximately the same
when calculated with the light-front OPE potential. However, the
high-momentum tail of the deuteron enhances the effects of the
potential's relativistic components. Since the breaking of rotational
invariance is a relativistic effect, this implies that the mass splitting
may be large for the deuteron calculated with the light-front potential.

To clearly understand how a large mass splitting might arise in the
light-front pion-only model, we take a step back and consider a scalar
version of the pion-only model, in which we assume that the pion has a
scalar coupling to the nucleon. This allows for a more direct comparison
with the Wick-Cutkosky model, since the main difference between the
scalar-pion-only potential and the Wick-Cutkosky potential (aside from
an isospin factor) is the pion-nucleon form factor. Since the
denominator of the form factor has the same form as the denominator of
the potential, the form factors do not significantly change the
rotational properties of the scalar-pion-only potential. Another
difference is that factors of $\overline{u}u$ appear in the numerator of
the scalar-pion-only potential, but the effect of those is small. 

The bound state calculations are performed using techniques detailed in
Refs.~\cite{Cooke:2001kz,Cooke:2001dc}, so we concern ourselves here
only with the results. 
In Table~\ref{pi:tab:checkscalarpionbe}, we show the binding energies
for deuterons calculated with the scalar-pion-only model. Two
pion-nucleon form factors are considered, the first one has
$\Lambda=1.0$~GeV, for which the coupling constant was fit to give the
correct binding energy for the non-relativistic potential, and the
second form factor uses $\Lambda=1.915$~GeV, which was fit to give the
correct binding energy for the light-front potential. Those form factors
are used to calculate the binding energies for the non-relativistic and
light-front potentials, as well as the instantaneous and retarded
potentials, which are relativistic and defined by analogy with
instantaneous and retarded potentials defined for the Wick-Cutkosky
model in Ref.~\cite{Cooke:2000ef}. We find that the binding energies for
all of the potentials have the same order of magnitude, and that the
light-front potentials have consistently lower binding energies, which
confirms the behavior observed in Ref.~\cite{Cooke:2000ef}. In addition,
the binding energies of the $m=0$ and $m=1$ light-front potentials are
essentially degenerate.  

Now we ready to consider the (pseudoscalar) pion-only model. The only
difference between this potential and the scalar-pion-only potential is
that the numerator contains factors of $\overline{u}i\gamma^5u$ instead
of $\overline{u}u$. Although this seems to be only a small change, it
has a large effect on the binding energies. The pseudoscalar coupling
generates a tensor force, which is more sensitive to the relativistic
components of the wave function than the scalar force. In general, this
means that the differences in binding energies obtained with different
potentials, such as those shown in Table~\ref{pi:tab:checkscalarpionbe},
will be larger for the pion-only potential. 

In Table~\ref{pi:tab:checkpionbe}, we show the
binding energies for deuterons computed  with the pion-only model. Two
pion-nucleon form factors are considered, the first one has
$\Lambda=1.01$~GeV, which was fit for the non-relativistic pion-only
model \cite{Friar:1984wi}, and the second form factor uses
$\Lambda=1.9$~GeV, which was fit to give the most reasonable binding
energy for the light-front potentials. We find that the binding energies
vary greatly depending on which potential is used. In fact, the
light-front potentials do not bind the deuteron with the first form
factor, and with the second form factor, the mass splitting between the
$m=0$ and $m=1$ states is very large. These facts indicate that the
deuteron wave functions are very sensitive to subtle changes in the
relativistic structure of the pion-only potentials.  

The reason for this mass splitting is that the OPE potential
breaks rotational invariance.
To reduce this splitting, higher-order potentials must be used.
Na\"{\i}vely, one might think that we can choose any set of
TPE graphs which is a truncation of a
rotationally-invariant infinite series of graphs. In particular, one
might think that using the TPE box graphs, which we denote as the
non-chiral-TPE (ncTPE) potential, would be adequate for our analysis of
rotational invariance. 

Another choice for the TPE potential is to take sum of the TPE
box graphs and the TPE contact graphs. To incorporate chiral symmetry as
accurately as possible without including the crossed TPE potentials, we
weight the contact vertex with a factor of
$\frac{M}{2(M+m_\pi)}$, as explained in earlier. Note that in
the sum, which we call the TPE potential, chiral symmetry provides a
cancelation between the box diagrams and the contact diagrams. This
indicates that the results obtained with the TPE potential will be more
stable than those obtained with the ncTPE potential.

To check this stability and the restoration of rotational invariance of
the pion-only model, we calculate the energy of the deuteron using the
OPE, OPE+TPE, and OPE+ncTPE potentials. We verify that the results are
independent of the choice of the pion-nucleon form factor by considering
three different choices for the form factor.

For the first pion-nucleon form factor, we choose $n=1$ and find that
$\Lambda=1.9$~GeV gives a reasonable range of binding energies. We use
the form factor to calculate the lowest energy bound state with
arbitrary total angular momentum. 

The results for the first pion-nucleon form factor are shown in
Table~\ref{pi:tab:Lam19n1}. 
Note that the results show that both the mass splitting and
the difference in the percentage of the D-state wave function decrease
when TPE diagrams are included. In addition, the wave functions are
almost completely in the $J=1$ state. As for the D-state probability, we
first note that it increases with the binding energy. It is consistent
(given the range of values it and the binding energy take) with the
value of 6\%
reported in Ref.~\cite{Friar:1984wi} for the FGP model.

Table~\ref{pi:tab:Lam19n1} also shows that numerically the states are
{\em almost} angular momentum eigenstates, since the percentage of
each state with $J=1$ is almost 100\%.
This happens even though the eigenstates of the light-front
Hamiltonian are not in principle eigenstates of the angular momentum.
However, this does not mean
that rotational invariance is unbroken. Rotations relate the different
$m$ states, and from the mass splittings, we see that the states do not
transform correctly.

Another thing to note in Table~\ref{pi:tab:Lam19n1} are the effects of
using the ncTPE potential. As mentioned earlier, this potential
is greater in magnitude than the TPE potential and should have a larger
effect on the binding energy. In fact, the effect is so large that it
serves to unbind the deuteron. (Strictly speaking, this indicates only
that the binding energy is very small or zero; the error in the binding
energy calculation increases as the binding energy approaches zero.)
Because the ncTPE potential has such a large effect by itself, it is
impossible to determine what effect it has on the rotational properties
of the state. Only the TPE potential can be used to analyze the
restoration of the state's rotational invariance.

To make sure that the results found are independent of the pion-nucleon
form factor, we performed the same calculation with $n=1$ and
$\Lambda=2.1$~GeV and with $n=2$ and $\Lambda=2.9$~GeV. The results
(shown in Ref.~\cite{Cooke:2001dc}) are qualitatively the same as those
in Table~\ref{pi:tab:Lam19n1}, demonstrating that these results are
robust.

\section{Conclusions} \label{ch:conclusions}

In this paper, a light-front pion-only model was used to investigate the
effects that relativity and chiral symmetry have on the deuteron. We
used OPE, OPE+TPE, and OPE+ncTPE potentials to calculate the binding
energy and wave function for the $m=0$ and $m=1$ states of the
deuteron. We find that the splitting between the $m=0$ and $m=1$ states
is smaller for the OPE+TPE potential as compared to what the OPE
potential gives. We also find that chiral symmetry must be included to
obtain sensible results when using two-pion-exchange potentials.

\begin{acknowledgments}
This work is supported in part by the U.S.~Dept.~of Energy under Grant
No.~DE-FG03-97ER4014.
We also thank Daniel Phillips for many useful discussions.
\end{acknowledgments}


\begin{table}
\caption{Binding energies for deuterons calculated with several
potentials and
two different pion-nucleon form factors for the scalar-pion-only
model. The $n$
parameter of the form factor is 1, the coupling constant is
$\frac{g^2}{4\pi}=0.424$, and the pion mass is used as the mass of the
exchanged meson. 
\label{pi:tab:checkscalarpionbe}}
\begin{center}
\begin{ruledtabular}\begin{tabular}{ccc}
Potential
& $\Lambda=1.0$~GeV  
& $\Lambda=1.915$~GeV 
\\ \hline
Non-relativistic   & 2.2236~MeV  & 4.2539~MeV \\ 
Instantaneous      & 2.1581~MeV  & 4.0862~MeV \\ 
Retarded           & 2.3111~MeV  & 4.5224~MeV \\ 
Light-front, $m=0$ & 1.2027~MeV  & 2.2296~MeV \\ 
Light-front, $m=1$ & 1.2027~MeV  & 2.2294~MeV \\ 
\end{tabular}\end{ruledtabular}
\end{center}
\end{table}

\begin{table}
\caption{Binding energies for deuterons calculated with several
potentials and
two different pion-nucleon form factors for the (pseudoscalar) pion-only
model. The
$n$ parameter of the form factor is 1, the coupling constant
is $\frac{g^2}{4\pi}=14.6$, and the pion mass is used as the mass of the
exchanged meson. The negative binding energy for the light-front
potentials in the first column indicates that the states are not bound. 
\label{pi:tab:checkpionbe}}
\begin{center}
\begin{ruledtabular}\begin{tabular}{ccc}
Potential
& $\Lambda=1.01$~GeV 
& $\Lambda=1.9$~GeV  
\\ \hline
Non-relativistic   &  2.2244~MeV  &  227.019~MeV  \\
Instantaneous      &  0.0146~MeV  &   28.192~MeV  \\
Retarded           &  1.3590~MeV  &  130.472~MeV  \\
Light-front, $m=0$ & -0.0269~MeV  &    0.788~MeV  \\
Light-front, $m=1$ & -0.0246~MeV  &    8.856~MeV  \\
\end{tabular}\end{ruledtabular} 
\end{center}
\end{table}

\begin{table*}
\caption{The values of the binding energy for the $m=0$ and $m=1$ states,
the difference of those binding energies ($\Delta$), percentage of
the wave function in the D state, and the percentage of the wave function
in the $J=1$ state for the $m=0$ and $m=1$ states for different
potentials.
The pion-nucleon form factor uses $n=1$ and $\Lambda=1.9$~GeV.
\label{pi:tab:Lam19n1}}
\begin{center}
\begin{ruledtabular}\begin{tabular}{cccccccc}
Potential &
\multicolumn{3}{c}{Binding Energy (MeV)} &
\multicolumn{2}{c}{D state (\%)} &
\multicolumn{2}{c}{$J=1$ (\%)} \\ \hline
& m=0 & m=1 & $\Delta$ & m=0 & m=1 & m=0 & m=1 \\ \hline
OPE       & 0.7884  & 8.8561 & -8.0677 & 4.09   &12.27   & 99.99    &99.78  \\
OPE+TPE   & 0.6845  & 2.5606 & -1.8761 & 3.98   & 8.08   & 99.98    &99.88  \\
OPE+ncTPE & -0.0107  & -0.0087 & -0.0020 & 0.15   & 0.66   &100.00    &100.00 \\
\end{tabular}\end{ruledtabular} 
\end{center}
\end{table*}


\end{document}